\documentstyle[amsfonts,aps]{revtex}
\def\a{\alpha}
\def\6{\partial}

\begin{document}
\draft
\title{Duality between coordinates and Dirac field}
\author{M. C. B. Abdalla, A. L. Gadelha and I. V. Vancea \thanks{%
On leave from Babes-Bolyai University of Cluj.}.}
\address{{\em Instituto de F\'{\i}sica Te\'{o}rica, Universidade Estadual
Paulista}\\ {\em Rua Pamplona 145, 01405-900, S\~{a}o Paulo, SP, Brazil}}
\maketitle

\begin{abstract}
The duality between the Cartesian coordinates on the Minkowski space-time and
the Dirac field is investigated. Two distinct possibilities to define this
duality are shown to exist. In both cases, the equations satisfied by
prepotentials are of second order.

\end{abstract}

\date{\today }

Recently, a duality between the Cartesian coordinates on the Minkowski
space-time and the solutions of Klein - Gordon equation could be derived\cite
{fm} by employing a method to invert the wave functions which was
previously used in the context of Seiberg-Witten theory \cite{sw}.

Several consequences of this duality were analyzed in subsequent
studies. Based on it, there was proposed an equivalence principle stating 
that 
all physical systems should be related by coordinate
transformation \cite{ep,lmm}. By applying the equivalence principle to the
phase-space in the Hamilton-Jacobi formulation it was discovered that the
energy quantization of the bound states and the tunneling effect can be
derived without any assumption about the probability interpretation of the
wave function \cite{ep}. In this way, a trajectory representation of the 
quantum
mechanics that was previously used in \cite{erf} can also be obtained.
The simplification of the coordinate-field duality at Planck scale and
for Heisenberg's uncertainty principle \cite{ps} was discussed in
\cite{fm,af}. In \cite{rc} it was analyzed a coordinate-free formulation of
gravity, while in' \cite{mv} the coordinate-field duality was extended to
curved space-time manifolds and in\cite{viv} there was proposed a local
formulation of gravity in terms of matter fields. (For recent attempts to
formulate the gravity and supergravity in terms of quantum objects, see
\cite{cr,viv1}.) Although the above mentioned results were obtained basically
from coordinate-field duality, one should note that, in the relativistic
case, this is constructed exclusively for the Klein-Gordon field. In reference
\cite{fm} it was suggested that the coordinate-field duality should hold
for the Dirac field, too. However, this suggestion was never realized in a 
precise way. In
this letter we follow \cite{fm} and concern ourselves with investigating the
coordinate-field duality in the case of the Dirac field. For definiteness, we
work in the four-dimensional Minkowski space-time with the metric signature
mostly minus and we assume that the field is a Dirac spinor. However, the
results can be straightforwardly generalized to different space-time
dimensions and to other types of spinors. Basic remark is that the linear
independent solutions of the Dirac equation for the field and its Dirac
conjugate can be factorized as follows 
\begin{eqnarray}
\psi _{\alpha }\left( k,s;x\right) &=&u_{\alpha }\left( k,s\right) \phi
\left( k,x\right), \qquad \widetilde{\psi }_{\alpha }\left( k,s;x\right)
=v_{\alpha }\left( k,s\right) \widetilde{\phi }\left( k,x\right),  \nonumber
\\
\overline{\psi }_{\alpha }\left( k,s;x\right) &=&\overline{u}_{\alpha
}\left( k,s\right) \widetilde{\phi }\left( k,x\right), \qquad \widetilde{%
\overline{\psi }}_{\alpha }\left( k,s;x\right) =\overline{v}_{\alpha }\left(
k,s\right) \phi \left( k,x\right) .  \label{sol}
\end{eqnarray}
Here, $u_{\alpha }\left( k,s\right) $ and $v_{\alpha }\left( k,s\right) $
are column vectors that span the space of $spin(1,3)$ spinors and they depend
only on the momentum and the spin of the field. The fields $\phi 
\left( k,x\right) $
and $\widetilde{\phi }\left( k,x\right) $ are linear independent solutions
of the Klein-Gordon equation which usually are taken to be wave functions. The
Dirac conjugation is denoted by a bar and the distinction between two
linearly independent solutions is made by a tilde. To simplify the relations,
we drop the indices $k$ and $s$ in what follows.

In order to determinate the coordinate-field duality for the Klein-Gordon 
field, 
one keeps a coordinate at a time as a variable, say $x^{\mu }$%
, and treat $x^{\nu }$ with $\mu \neq \nu $ as parameters \cite{fm}. The
corresponding solutions are labelled with an upper index ($\mu $) which is
a dumb index, e.g. $\phi ^{\left( \mu \right) }$. Then a prepotential ${\cal %
F}_{\phi }^{\left( \mu \right) }\left[ \phi ^{\left( \mu \right) }\right] $
is introduced for each pair of linearly independent solution by \cite
{fm,mm,bm} 
\begin{equation}
\widetilde{\phi }^{\left( \mu \right) }\equiv \frac{{\cal F}_{\phi }^{\left(
\mu \right) }\left[ \phi ^{\left( \mu \right) }\right] }{\partial \phi
^{\left( \mu \right) }}.  \label{prepsacal}
\end{equation}
In the case of the Dirac field, we can follow the same procedure. However, 
since
the dependence on $x$ of $\psi _{\alpha }$ and $\overline{\psi }_{\alpha }$
is carried entirely by $\phi $ and $\widetilde{\phi }$ according to (\ref
{sol}), there is an ambiguity which in the case of the Klein-Gordon field does
not appear: namely, one can use either $\widetilde{\psi }_{\alpha }$ or $%
\overline{\psi }_{\alpha }$ to define the prepotential for $\psi _{\alpha }$%
. This ambiguity is due to the fact that the second-order Klein-Gordon
equation has been split into two first-order differential equation. The
second possibility which implies mixing the solutions of the Dirac equation for
the field and its conjugate is treated as the first one from the point
of view of the second-order differential equation. We analyze both cases.

In the first case we define the prepotential of the Dirac field by the
following relation 
\begin{equation}
\widetilde{\psi }_{\alpha }^{\left( \mu \right) }\equiv \frac{\partial {\cal %
F}_{\alpha \psi }^{\left( \mu \right) }\left[ \psi _{\alpha }^{\left( \mu
\right) }\right] }{\partial \psi _{\alpha }^{\left( \mu \right) }},
\label{prepfirst}
\end{equation}
for $\mu =0,1,2,3,$ where $\widetilde{\psi }_{\alpha }^{\left( \mu \right) }$
are the solutions of the Dirac equation ``along the direction $x^{\mu }$'',
i.e. with $x^{\nu }$= const. for $\nu \neq \mu $. It is easy to see that $%
\psi _{\alpha }^{\left( \mu \right) }$ factorizes as in (\ref{sol}) with $%
\phi ^{\left( \mu \right) }$ satisfying the corresponding Klein-Gordon
equation \cite{fm}. In order to express $x^{\mu }$ as a function of $\psi
_{\alpha }^{\left( \mu \right) }$ and ${\cal F}_{\alpha \psi }^{\left( \mu
\right) }$, one has to derive the prepotential with respect to the
space-time coordinate \cite{fm}. However, the usual derivation does not make
sense since it includes the product $v_{\alpha }u_{\alpha }$ which is not
defined for two column vectors. To makeshift around this difficulty we
define the derivative of the prepotential through a tensor product by 
\begin{equation}
\partial _{\mu }{\cal F}_{\alpha \psi }^{\left( \mu \right) }=\frac{\partial 
{\cal F}_{\alpha \psi }^{\left( \mu \right) }}{\partial \psi _{\alpha}^
{\left( \mu \right) }}\otimes \frac{\partial \psi _{\alpha }^{\left( \mu
\right) }}{\partial x^{\mu }}.  \label{tensor}
\end{equation}
The definition (\ref{tensor}) takes into account the internal structure of
the Dirac field. The tensor product $v_{\alpha }\otimes u_{\alpha }$ can be
identified with an invertible $4\times 4$ matrix. By a straightforward 
computation,
one can verify that the duality between the coordinates and fields is given
by the following relation 
\begin{equation}
\frac{\sqrt{2m}}{\hbar } X_{\a \a }^{\mu }=\frac{1}{2}
\frac{\6{\cal F}_{\a \psi }^{\left( \mu \right) }}
{\6\psi _{\a}^{\left( \mu \right) }}
\otimes \psi _{\a }^{\left( \mu \right) }
-
{\cal F}_{\alpha \psi }^{\left( \mu \right) }+
C_{\alpha }^{\left( \mu \right) },
\label{dualone}
\end{equation}
for $\mu =0,1,2,3$. Here 
$X_{\alpha \alpha }^{\mu }=O_{\alpha \alpha }x^{\mu}\equiv v_{\alpha }
\otimes u_{\alpha }x^{\mu }$. The arbitrary function $%
C_{\alpha }^{\left( \mu \right) }$ does not depend on $x^{\mu }$ but depends
on the parameters $x^{\nu }$ and on the momentum and spin of the field $\psi
_{\alpha }^{\left( \mu \right) }$.

In order to completely determine the duality (\ref{dualone}) one has to find
out the differential equation satisfied by the prepotential. The method for
doing this was described in\cite{mm,fm}. By applying it in the present case, 
one can show that ${\cal F%
}_{\alpha \psi }^{\left( \mu \right) }$ satisfies the following differential
equation 
\begin{equation}
i\frac{\sqrt{8m}}{\hbar }\gamma ^{\mu }O_{\alpha \alpha }\delta _{\mu }^{2}%
{\cal F}_{\alpha \psi }^{\left( \mu \right) }+\left[ \widetilde{V}_{\alpha
}^{\left( \mu \right) }-m\right] \delta _{\mu }{\cal F}_{\alpha \psi
}^{\left( \mu \right) }\left( \delta _{\mu }^{2}{\cal F}_{\alpha \psi
}^{\left( \mu \right) }\otimes \psi _{\alpha }^{\left( \mu \right) }-\delta
_{\mu }{\cal F}_{\alpha \psi }^{\left( \mu \right) }\right) =0,
\label{egprepone}
\end{equation}
for $\mu =0,1,2,3,$ where $\delta _{\mu }=\partial /\partial \psi _{\alpha
}^{\left( \mu \right) }$. The potential $\widetilde{V}_{\alpha }^{\left(
\mu \right) }$ in (\ref{egprepone}) is given by 
\begin{equation}
\widetilde{V}_{\alpha }^{\left( \mu \right) }=\left. \left[
i\sum\limits_{\upsilon \neq \mu }\gamma ^{\nu }\partial _{\nu }\widetilde{%
\psi }_{\alpha }^{\left( \mu \right) }\cdot \widetilde{\overline{\psi }}%
_{\alpha }^{\left( \mu \right) }\right] \right| _{x^{\nu }=ct;\nu \neq \mu }.
\label{potone}
\end{equation}
Some comments are in order now. From (\ref{dualone}) we see that the
coordinate-field duality gives different ``representations'' of the
coordinate $x^{\mu }$, corresponding to different solutions of the Dirac 
equation labelled by $%
\alpha $. This is to be expected, since the Dirac field has an inner structure 
manifest
 in the factorization (\ref{sol}). If the matrix $O_{\alpha \alpha}$
is invertible we can express the real coordinates as functions of the
spinor fields. In this case, we expect that the image $x^{\mu
} $ $\left[ \widetilde{\psi }_{\alpha }^{\left( \mu \right) }\right] $ be an
unique real number. If this is the case, on has to impose a constraint on
the prepotentials which for this system is given by the following relation
\begin{equation}
\left( O_{11}\right) ^{-1}\left[ \frac{1}{2}\frac{\partial {\cal F}_{1\psi
}^{\left( \mu \right) }}{\partial \psi _{1}^{\left( \mu \right) }}\otimes
\psi _{i}^{\left( \mu \right) }-{\cal F}_{1\psi }^{\left( \mu \right)
}+C_{1}^{\left( \mu \right) }\right] =\left( O_{22}\right) ^{-1}\left[ 
\frac{1}{2}\frac{\partial {\cal F}_{2\psi }^{\left( \mu \right) }}
{\partial \psi
_{2}^{\left( \mu \right) }}\otimes \psi _{2}^{\left( \mu \right) }-{\cal F}
_{2\psi }^{\left( \mu \right) }+C_{2}^{\left( \mu \right) }\right] .
\label{coustr}
\end{equation}
Also, due to the fact that the field satisfies a first-order differential
equation, the prepotential satisfy a second-order differential equation (\ref
{egprepone}). We recall that in the Klein-Gordon case the corresponding
equation is a third-order one. Another difference from the Klein-Gordon field 
is
that the potential function (\ref{potone}) involves, beside the contribution
from the parameters $x^{\nu }$, a solution of the conjugate Dirac equation,
which shows that both the Dirac field and the conjugate Dirac field should
be considered in the coordinate-field duality.
 Note than, when the matrix
$O_{\alpha \alpha}$ is not invertible, although the real coordinates should be 
the same in all representations, the relation (\ref{coustr}) can not be imposed 
as a constraint.

In the second case one defines the prepotential through the relation 
\begin{equation}
\overline{\psi }_{\alpha }^{\left( \mu \right) }\equiv \frac{\partial 
{\cal F%
}_{\alpha \psi }^{\left( \mu \right) }\left[ \psi _{\alpha }^{\left( \mu
\right) }\right] }{\partial \psi _{\alpha }^{\left( \mu \right) }}.
\label{prepsecond}
\end{equation}
The duality relation obtained from (\ref{prepsecond}) is given by 
\begin{eqnarray}
\frac{\sqrt{2m}}{\hbar }x^{\mu } &=&\frac{1}{2}\frac{\partial {\cal F}%
_{\alpha \psi }^{\left( \mu \right) }}{\partial \psi _{\alpha }^{\left( \mu
\right) }}\psi _{\alpha }^{\left( \mu \right) }-{\cal F}_{\alpha \psi
}^{\left( \mu \right) }+C_{\alpha }^{\left( \mu \right) }  \nonumber \\
&=&\frac{1}{2}\frac{\partial {\cal F}_{\phi }^{\left( \mu \right) }}{%
\partial \phi ^{\left( \mu \right) }}\phi ^{\left( \mu \right) }-{\cal F}%
_{\phi }^{\left( \mu \right) },  \label{dualsecond}
\end{eqnarray}
for $\mu =0,1,2,3$. In deriving the second equality, we have used the fact
that the prepotential depends only on a solution of the Dirac equation with
given momentum and spin and we have used the normalization of $\overline{u}%
_{\alpha }\left( k,s\right) $ and $u_{\alpha }\left( k,s\right) $  to
one. The relation (\ref{dualsecond}) shows that in the second case the
``representation'' of coordinate $x^{\mu }$ in terms of the Dirac field is
unique. In order to find the equation for the prepotentials, we have to assume
that the matrix $O_{\alpha \alpha }$ is invertible. This allows us  to write a 
potential functional on each directions. Then the prepotential defined in 
(\ref{prepsecond}) satisfies the
 following differential equation 
\begin{equation}
i\frac{\sqrt{8m}}{\hbar }\delta _{\mu }^{2}{\cal F}_{\alpha \psi }^{\left(
\mu \right) }\gamma ^{\mu }+\left[ \overline{V}_{\alpha }^{\left( \mu
\right) }-m\right] \delta _{\mu }{\cal F}_{\alpha \psi }^{\left( \mu \right)
}\left( \delta _{\mu }^{2}{\cal F}_{\alpha \psi }^{\left( \mu \right) }\psi
_{\alpha }^{\left( \mu \right) }-\delta _{\mu }{\cal F}_{\alpha \psi
}^{\left( \mu \right) }\right) =0  \label{egpreptwo}
\end{equation}
where the potential $\overline{V}_{\alpha }^{\left( \mu \right) }$ has the
following form 
\begin{equation}
\overline{V}_{\alpha }^{\left( \mu \right) }=\left. \left[
i\sum\limits_{\upsilon \neq \mu }\gamma ^{\nu }\partial _{\nu }\overline{%
\psi }_{\alpha }^{\left( \mu \right) }O_{\alpha \alpha }^{-1}\psi _{\alpha
}^{\left( \mu \right) }\right] \right| _{x^{\nu }=ct;\nu \neq \mu },
\label{pottwo}
\end{equation}
where $O_{\alpha \alpha }=u_{\alpha }\otimes \overline{u}_{\alpha }$. Note
that although the tensor product disappeared from the duality relation, it
still remains in the equation for ${\cal F}_{\alpha \psi }^{\left( \mu
\right) }$ as in the first case. Due to this fact, the potential (\ref
{pottwo}) acts as a tensor product on the next term in (\ref{egpreptwo}).
However, one has to note that when the matrix $O_{\alpha \alpha }$ is not
invertible, which is the case for most of the representations of the spinor
field, the explicit form of the potential (\ref{pottwo}) is unknown and
the resolution of the second case along the line of \cite{fm} is problematic
and might very well not exist. 

The relations (\ref{dualone}), (\ref{egprepone}) and (\ref{dualsecond}), (%
\ref{egpreptwo}) describe the coordinate-field duality and the equations
of prepotentials for the Dirac field in the two cases allowed by the 
Klein-Gordon equation. When one takes for $\phi \left( k,x\right) $ and $%
\widetilde{\phi }\left( k,x\right) $ the corresponding wave function, the
entire dependence of $x^{\mu }$ on the Dirac fields is concentrated in $%
{\cal F}_{\alpha \psi }^{\left( \mu \right) }.$ Note that in the quantum
case, the factors $u_{\alpha }$ and $v_{\alpha }$ include anticommuting
creation and annihilation operators. Also, since the prepotential are
functionals of fields, they can be in principle either anticommuting or
commuting. In both cases the equations (\ref{egprepone}) and 
(\ref{egpreptwo}) become of first degree, since the prepotentials are 
polynomials of rank one in the fields. Then (\ref{egprepone}) and 
(\ref{egpreptwo}) can be easily solved. Two cases are to be
considered. In the first one $V_{\alpha }^{\left( \mu \right) }=m$ and the
prepotential can be an arbitrary polynomial of rank one in the Dirac field.
 In the
second case $V_{\alpha }^{\left( \mu \right) }\neq m$ and the prepotential
is an arbitrary constant functional.

Finally let us discuss the symmetry of the coordinate-field duality. In 
\cite{fm} it was shown that the duality between the coordinate and the 
wave-function in the quantum
mechanics obeys the modular symmetry $SL\left( 2,{\Bbb C}\right) $ of the
solution of the Schr\"{o}dinger equation. In particular, under an 
$SL\left( 2,{\Bbb C}\right) $ transformation of the linearly independent 
solutions, the
prepotentials transform as (16) of \cite{fm}. In the case of the 
Dirac field we
have the same situation. Indeed, if we consider the coordinate-field
duality defined by (\ref{dualsecond}) on can reproduce (16) of \cite{fm}
if we transform the linearly independent solution 
($\psi ,\widetilde{\psi }$) by $\!\left( \begin{array}{cc}
A & B \\ 
C & D
\end{array}
\right) \in SL\left( 2,{\Bbb C}\right) $ (for the sake of clarity, we 
omit all
 the
indices). In this case the problem reduces to the Klein-Gordon
problem as can be seen from the second equality from (\ref{dualsecond}). For
the first coordinate-field duality given by (\ref{dualone}) one has to
introduce the circle product between a line vector and a column
vector with Dirac field components $\left( 
\begin{array}{cc}
\mu _{1} & \mu _{2}
\end{array}
\right) $ and $\left( \begin{array}{c}\nu _{1} \\ 
\nu _{2}
\end{array}
\right) $ given by the following relation:
\begin{equation}
\left( 
\begin{array}{cc}
\mu _{1} & \mu _{2}
\end{array}
\right) \circ \left( 
\begin{array}{c}
\nu _{1} \\ 
\nu _{2}
\end{array}
\right) =\mu _{1}\otimes \nu _{1}+\mu _{2}\otimes \nu _{2}.  \label{dot}
\end{equation}
The properties of (\ref{dot}) are given by properties of the tensor product.
The product (\ref{dot}) is allowed because there is
an ambiguity in defining a product between two vectors which have as
components others vectors. Using (\ref{dot}), is easy to verify that under 
$SL\left( 2,{\Bbb C}\right) $
the prepotential ${\cal F}$ transforms as 
\begin{equation}
\delta {\cal F}=\frac{1}{2}{\cal X}\circ \left[ G^{T}\left( 
\begin{array}{cc}
0 & 1 \\ 
1 & 0
\end{array}
\right) G-\left( 
\begin{array}{cc}
0 & 1 \\ 
1 & 0
\end{array}
\right) \right] {\cal X},  \label{transf}
\end{equation}
where ${\cal X}$ is the two components column vector $\left( 
\begin{array}{c}
\psi \\ 
\widetilde{\psi }
\end{array}
\right) $.

In conclusion, the coordinate-field duality for the Dirac field has
two possible forms given by (\ref{dualone}) and (\ref{dualsecond}),
respectively. Both of those forms are compatible with the Klein Gordon
equation. However in the first case the representation of the Cartesian
coordinates through the Dirac field is degenerate, while in the second case
it is equivalent with the one obtained from the duality for Klein-Gordon
field. In both cases, the prepotentials transform under 
$SL\left( 2,{\Bbb C}\right) $
group as in (\ref{transf}) with the only 
difference that in the
second case the circle product should be replaced by the usual product.
It is important to note that we have used 
the fact that the matrix $O_{\alpha \alpha}$ is invertible in deriving the
equations satisfied by prepotentials. This matrix depends crucially on 
the linear independent basis $u_{\alpha}$ and $v_{\alpha}$
as well as on the mapping of the tensor product of these vectors to quadratic 
matrices. An unappropriate choice of any of these two elements can
lead to an univertible matrix. Nevertheless, one can always find a suitable 
basis and define the mapping in such of way that the inverse of
$O_{\alpha \alpha}$ exists.

The fact that 
there are several ``representations'' of coordinates in terms of
fields may bring some simplifications in the study of coordinate-field
duality for supersymmetric systems.

We want to thank M. A. de Andrade, J. A. Helayel-Neto and A. I. 
Shimabukuro for useful discussions.
A.L.G. acknowledgs V. S. Tim\'oteo for discussions and CAPES for his 
fellowship. The work of I.V.V. was supported by a FAPESP postdoc fellowship.

\end{document}